\documentclass[conference]{IEEEtran}

\usepackage{cite}
\usepackage{amsmath,amssymb}
\usepackage{graphicx}
\usepackage{booktabs}
\usepackage{url}
\usepackage{hyperref}
\usepackage{microtype}
\usepackage{xcolor}
\usepackage{enumitem}

\setlist{noitemsep, topsep=2pt, parsep=0pt}

\newcommand{\AIModel}{OpenAI GPT-4.1}

\title{The Brand War: A Gamified AI-Feedback System for Time-Limited EFL Writing}

\author{
    \IEEEauthorblockN{
        Jing-Yuan Huang\IEEEauthorrefmark{1},
        Vivien Lin\IEEEauthorrefmark{2},
        Yujong Park\IEEEauthorrefmark{3},
        Yi Miao\IEEEauthorrefmark{4},
        Yun-Hua Hsiao\IEEEauthorrefmark{2},
        Michael Pin-Chuan Lin\IEEEauthorrefmark{5}\IEEEauthorrefmark{1},\\
        Daniel Chang\IEEEauthorrefmark{6}\IEEEauthorrefmark{3},
        Seong Min Park\IEEEauthorrefmark{4},
        Marco Ho\IEEEauthorrefmark{7},
        Michael S. Hsiao\IEEEauthorrefmark{8}, and
        Jeeho Ryoo\IEEEauthorrefmark{4}
    }
    \IEEEauthorblockA{\IEEEauthorrefmark{1}Simon Fraser University, Canada}
    \IEEEauthorblockA{\IEEEauthorrefmark{2}National Changhua University of Education, Taiwan}
    \IEEEauthorblockA{\IEEEauthorrefmark{3}Sungkyunkwan University, Korea (South)}
    \IEEEauthorblockA{\IEEEauthorrefmark{4}Fairleigh Dickinson University, Canada}
    \IEEEauthorblockA{\IEEEauthorrefmark{5}Mount Saint Vincent University, Canada}
    \IEEEauthorblockA{\IEEEauthorrefmark{6}Athabasca University, Canada}
    \IEEEauthorblockA{\IEEEauthorrefmark{7}British Columbia Institute of Technology, Canada}
    \IEEEauthorblockA{\IEEEauthorrefmark{8}Virginia Tech, USA}
}

\begin{document}

\bstctlcite{IEEEexample:BSTcontrol}

\maketitle

\begin{abstract}
Writing is a cognitively demanding and anxiety-provoking skill for English as a Foreign Language (EFL) learners, particularly under time-limited conditions. This paper presents \textit{The Brand War}, a web-based gamified writing application that combines competitive game mechanics with iterative \AIModel-powered formative feedback for undergraduate EFL learners in a timed narrative writing task. The application situates students as marketing interns competing for a job offer, using review passes to receive AI feedback, attack opponents, or shield their own passes while drafting a 500-word brand story. We conducted an exploratory single-session classroom study with $N = 29$ university EFL students in Taiwan to examine how students engaged with the system, whether iterative AI feedback improved writing performance across revision attempts, and how AI and human scores related to overall outcomes. Students wrote their brand story within 60 minutes, using up to five AI feedback passes before a final human-graded submission. Most students ($65.5\%$) used the AI feedback system, and within-student AI scores improved modestly across revision attempts ($M = +3.7$ points, $SD = 7.4$), with larger gains among students completing more review cycles and scores significantly higher on the final review than the first among those who completed multiple cycles ($p = .032$). AI-assessed and human final scores showed strong convergent validity ($r = 0.722$, $p < 0.001$), and students who used AI feedback scored descriptively, though not significantly, higher than non-users. Students maintained a high mean focus ratio of $82.4\%$, and competitive mechanics were used sparingly, suggesting most students prioritized writing over social interference even when such options were available. Findings suggest that embedding iterative AI scoring within a competitive game context is feasible and may scaffold writing improvement, with implications for EFL writing pedagogy and AI-mediated gamified learning design.

\end{abstract}

\begin{IEEEkeywords}
gamification, EFL writing, AI feedback, formative assessment, game-based learning, self-regulated learning
\end{IEEEkeywords}

\section{Introduction}
Writing in a second or foreign language is cognitively demanding and anxiety-inducing \cite{guvendir2023writing} because writers juggle meaning-making and language control simultaneously, and the gap between what they intend to express and what they can express in the second language produces both cognitive strain and self-doubt. For EFL (English as a Foreign Language) writers, the challenge is compounded when writing is time-constrained, such as in course exams, proficiency assessments, or professional workplace tasks. Despite the prevalence of timed writing in real-world contexts, instructional approaches that effectively support EFL writers under such conditions remain underdeveloped.

Two independent lines of research offer complementary solutions. First, \textit{gamification}, the application of game design elements in non-game contexts \cite{deterding2011game}, has demonstrated positive effects on learner motivation and engagement in EFL writing tasks \cite{zhihao2022impact,guo2024applying}. Competitive game mechanics, narrative storylines, and reward structures can transform writing from a solitary, anxiety-laden exercise into an engaging, goal-directed activity. Second, AI-generated \textit{formative feedback} has emerged as a scalable tool for assessing student writing, with studies showing that while human evaluators retain advantages in nuanced prioritization and affective tone, AI scoring demonstrates convergent validity with expert raters on criteria-based dimensions \cite{steiss2024comparing}. The provision of rapid, rubric-aligned feedback enables iterative revision cycles that conventional classroom instruction rarely affords.

Despite these parallel advances, few systems have attempted to unify both approaches in a single learning environment. Most gamified writing studies use non-adaptive mechanics (e.g., leaderboards and badges) without coupling them to substantive writing feedback \cite{guo2024applying}. Conversely, AI feedback systems are typically deployed in standalone tools that lack the motivational architecture of a game. The combination, where game mechanics create urgency and investment while AI feedback guides improvement within that game, represents a largely unexplored design space.

This paper introduces \textit{The Brand War}, a web-based application designed to address this gap. In The Brand War, undergraduate EFL students assume the role of marketing interns competing for a job offer, writing a 500-word brand story under a 60-minute time limit. Students are allocated five AI feedback passes, each enabling one cycle of \AIModel-powered scoring and evaluation of students' current draft. Competitive mechanics, including the ability to attack opponents' passes and deploy shields, add an optional layer of social competition.

We conducted an exploratory single-session classroom study with $N = 29$ university EFL students in Taiwan to investigate three research questions.
\begin{itemize}
    \item \textbf{RQ1.} How do students engage with a gamified AI-feedback writing system during a time-limited writing task?
    \item \textbf{RQ2.} Does iterative AI feedback improve students' writing performance across successive revision attempts?
    \item \textbf{RQ3.} How do convergent AI and human scores, and the use of competitive mechanics, relate to overall writing outcomes?
\end{itemize}
The remainder of this paper reviews relevant literature (Section~II), describes the system design (Section~III), presents the study methodology (Section~IV), reports results (Section~V), and discusses implications and limitations (Sections~VI--VII).

\section{Literature Review}
\subsection{Gamification in EFL Writing}

Deterding et al. \cite{deterding2011game} define gamification as ``the use of game design elements in non-game contexts,'' distinguishing it from full game-based learning by its application of individual components, such as points, levels, and leaderboards, to existing activities. Zhihao and Zhonggen \cite{zhihao2022impact} showed that gamified elements improve learners' engagement and writing quality in time-limited tasks, though over-gamification can diminish motivation once reward saturation is reached. A scoping review by Guo et al. \cite{guo2024applying} synthesized 22 studies on game-related writing methods, finding digital games the most common format and competitive mechanics and storylines effective for engagement, but noting a scarcity of designs that couple mechanics to substantive feedback. Lin et al. \cite{lin2026effectiveness} similarly found that a tangible language game combining competition and collaboration (``coopetition'') was accepted across proficiency bands, with weaker students gaining the most, suggesting competitive design is a tunable variable rather than a binary choice.

\vspace{-2mm}
\subsection{Game-Based Learning and Narrative Writing}

Beyond gamification, full game-based learning (GBL) environments that situate students within immersive storylines have shown promising outcomes for writing motivation and performance. Narrative framing transforms writing into purposeful, audience-directed communication, reducing the abstraction that often contributes to EFL writing anxiety \cite{xiao2025using}, consistent with frameworks that frame writing within professional or scenario-driven narratives to reduce anxiety and L2 fatigue \cite{lin2018flipped}. Narrative-driven GBL also offsets disengagement during extended writing practice, with enjoyment predicting sustained motivation and skill gains across L1 and L2 cohorts \cite{allen2014l2}. Xiao et al. \cite{xiao2025using} integrated ChatGPT-powered non-player characters (NPCs) into storyline-driven GBL for argumentative EFL writing, finding higher intrinsic motivation, situational interest, and essay quality without increased cognitive load, an approach that informed The Brand War's professional narrative scenario, with AI extended here from an interactional NPC role to a formative scoring agent.

\vspace{-2mm}
\subsection{AI Feedback in EFL Writing}
Theoretical models have proposed how GenAI can support self-regulated learning and writing instruction \cite{chang2023educational,lin2023chat}, and the use of LLMs for writing assessment has accelerated rapidly. Steiss et al. \cite{steiss2024comparing} compared ChatGPT-generated feedback with human expert feedback, finding AI feedback approached human quality on surface-level linguistic features, though human raters retained an edge in nuanced, holistic judgment. AI-based scoring is nonetheless cost-effective and immediate, well-suited to time-constrained classrooms where human feedback cycles are impractical. Huang and Chen \cite{huang2025design} proposed the Gamified Self-Regulated English Learning (G-SERL) system, combining ChatGPT assistance with gamified reinforcement and self-regulation nudges, reporting gains in post-test writing performance relative to a ChatGPT-only control.

\vspace{-2mm}
\subsection{Theoretical Framework}

Three complementary theories ground the design and interpretation of The Brand War. \textit{Self-Determination Theory} (SDT) \cite{deci2000and} posits that intrinsic motivation is sustained when three basic psychological needs, autonomy, competence, and relatedness, are met. The game's competitive scenario supports \textit{relatedness}, the iterative AI scoring system supports \textit{competence} through immediate, specific feedback, and the choice to use passes offensively or constructively supports \textit{autonomy}. \textit{Self-Regulated Learning} (SRL) theory \cite{zimmerman2000attaining} frames learning as a cyclical process of planning, monitoring, and reflection, which the AI feedback pass mechanic operationalizes as students draft, receive evaluative information, set micro-goals, and revise. Finally, Vygotsky's \cite{vygotsky1978mind} \textit{Zone of Proximal Development} (ZPD) offers a scaffold-oriented lens, in which the \AIModel{} scoring model acts as a ``more knowledgeable other'' that provides performance signals students cannot easily generate for themselves.

\subsection{Research Gap}

Existing research has separately shown that gamification supports EFL writing engagement, that narrative contexts reduce writing anxiety, and that AI feedback can reliably assess and improve student writing. However, no system \textit{integrates} competitive game mechanics, a storyline-driven writing task, and iterative AI formative scoring within a single real-classroom deployment. The present study addresses this gap with an exploratory dataset from a live classroom implementation in Taiwan.

\section{System Design: The Brand War}
\subsection{Learning Context and Task}

The Brand War writing app was designed for an undergraduate mandatory EFL writing course at a university in Taiwan. The target task was a 500-word narrative brand story that students were required to write within 60 minutes. The brand story writing task required the participants to incorporate five narrative elements covered in the preceding instructional unit, namely \textit{setting}, \textit{theme}, \textit{mood}, \textit{character}, and \textit{plot}. This task design reflects authentic genre demands, since brand storytelling is increasingly recognized as a professional competency in marketing contexts, and provides a motivational anchor aligned with the game scenario.

\subsection{Game Scenario and Storyline}

To frame the writing task within a competitive narrative, The Brand War situates students as candidates competing for a summer internship at ``Global Marketing, Inc.,'' a fictional top-tier marketing firm at Taipei 101. The instructor introduces the scenario in the following terms.

\begin{quote}
\textit{``This summer, you will be looking for summer internships at Global Marketing, Inc. located in Taipei 101 to experience the real world of marketing. In the Brand War, you need to get to the top by getting a job offer. You must finish writing a brand story to bring to your interview.''}
\end{quote}
This storyline grounds the abstract writing task in a concrete professional situation, creates stakes through competition, and aligns with students' future career aspirations, all of which are consistent with SDT's emphasis on \textit{relevance} as a driver of autonomous motivation \cite{deci2000and}.

\subsection{Game Mechanics}

The application implements three distinct mechanic layers, summarized in Table~\ref{tab:mechanics}.
\begin{table}[tbp]
\caption{Summary of Game Mechanics in The Brand War}
\label{tab:mechanics}
\centering
\begin{tabular}{p{2cm} p{5.5cm}}
\toprule
\textbf{Mechanic} & \textbf{Description} \\
\midrule
Review Pass & Each student begins with 5 passes. Spending one pass submits the current draft for \AIModel{} scoring and feedback. Up to 5 revision cycles are possible. \\
\addlinespace
Attack & A student may spend a pass to destroy one of an opponent's passes, reducing the target's remaining feedback opportunities. \\
\addlinespace
Shield & Each student holds one shield. Activating it protects a pass from the next incoming attack. \\
\addlinespace
Final Submit & At any point before the time limit, a student submits their current draft as the final essay for human grading. \\
\bottomrule
\end{tabular}
\end{table}
The pass economy creates a meaningful trade-off. Using a pass for AI feedback improves the student's own draft, while using it to attack an opponent reduces a competitor's revision capacity. This trade-off is intended to sustain competitive tension without making competition the dominant strategy, because writing quality, after all, determines the job offer. Additionally, the system includes an attack validation rule requiring targets to retain at least one remaining pass if no AI feedback passes have been used, guaranteeing all participants at least one iterative AI feedback cycle regardless of competitive interference. Importantly, the attack mechanic only affects optional AI feedback opportunities and does not influence human-assigned grades or assessment outcomes.

\subsection{AI Feedback and Scoring System}

When a student submits a draft using a review pass, the essay is evaluated by \AIModel{} using a rubric-aligned prompt covering three dimensions, namely \textit{Content} (relevance, narrative richness, use of the five narrative elements), \textit{Mechanics} (grammar, spelling, punctuation, sentence structure), and \textit{Structure} (organization, coherence, introduction--body--conclusion flow). Each dimension is scored on a 0--100 scale, and a composite average is computed and displayed to the student alongside the dimensional breakdown. The rubric was developed by the course instructor and was embedded directly in the \AIModel{} prompt to ensure scoring alignment with classroom expectations. Students receive their scores immediately upon submission and may revise their essay before using another pass.

It is important to note that this AI feedback is best characterized as \textit{evaluative scaffolding} rather than full formative feedback. Students receive scores and dimensional breakdowns but not explicit revision suggestions. This design choice preserves writing agency and prevents over-reliance on AI-generated text, while still providing the diagnostic signal students need to guide self-directed revision, consistent with SRL theory's emphasis on self-monitoring and self-evaluation \cite{zimmerman2000attaining}.

\subsection{Engagement Tracking}

\begin{figure}[tbp]
    \centering
    \includegraphics[width=\columnwidth]{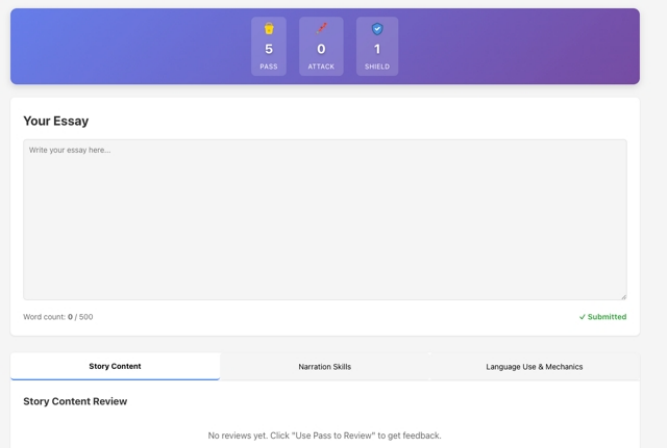}
    \caption{The Brand War application interface.}
    \label{fig:system}
\end{figure}

The application logs all user interactions, including window focus and blur events (via the browser Page Visibility API), submission timestamps, and attack/shield activations. From these logs, the platform computes student-level engagement metrics including \textit{focus ratio} (proportion of total task time during which the essay input text box window was active), \textit{number of focus sessions}, \textit{total task duration}, and \textit{passes used count}. Fig.~\ref{fig:system} shows the student-facing interface, with a resource bar tracking remaining passes, attacks, and shields, a live word-count essay editor targeting the 500-word limit, and a tabbed review panel that returns dimensional AI feedback once a pass is spent, keeping resource state, drafting progress, and feedback visible together.

\section{Methodology}
\subsection{Participants}

Participants were $N = 29$ university students enrolled in an elective EFL writing course at a university in Taiwan. The class represented a mix of majors. Two students did not submit a final essay and are excluded from human-score analyses, yielding $n = 27$ for those comparisons. One student had no focus-event logs and is excluded from focus-ratio analyses ($n = 28$). Demographics were reported based on the 27 participants who completed the final essay and post-survey (21 second-year, 2 fourth-year or above undergraduate students, and 2 graduate students).
Participants reported varied English proficiency based on self-reported standardized test results. TOEIC was the most common measure ($M = 824.31$, $n = 16$), and GEPT respondents generally reported intermediate/high-intermediate proficiency or above ($n = 12$). Data were collected under institutional research protocols during regular class hours as part of the scheduled course activity.

\vspace{-1mm}
\subsection{Procedure}

The session was conducted on March 17, 2026, and lasted approximately 80 minutes, in five phases.

\begin{enumerate}
    \item \textbf{Game Introduction} (${\sim}5$ min). The instructor introduced the scenario and rules (passes, attacks, shield, time limit).
    \item \textbf{Platform Onboarding} (${\sim}5$ min). Students logged in and confirmed they understood the prompt and rubric.
    \item \textbf{Writing and Revision} (${\sim}60$ min). Students composed their brand stories, optionally spending review passes for \AIModel{} scoring and revising accordingly, or using passes to attack opponents.
    \item \textbf{Final Submission}. Students clicked ``Submit Final Essay'' when satisfied or when time expired. Essays were graded by the instructor using the same three-dimensional rubric (Content, Mechanics, and Structure, each on a 0--100 scale) used by the AI system.
    \item \textbf{Debrief} (${\sim}5$ min). Brief class reflection on the writing process and game experience.
\end{enumerate}

\vspace{-1mm}
\subsection{Data Sources}

Two datasets were extracted from the platform logs. The \textbf{review-level data} ($n_{\text{reviews}} = 43$ submissions from 19 students) record student ID, attempt number (1--5), and Content, Mechanics, and Structure scores plus the composite AI average. The \textbf{student-level data} ($N = 29$) record focus ratio, total task duration, number of focus sessions, passes used, first and last AI review scores, AI score gain, attack count, shield use, and human final score.

\begin{figure}[tbp]
    \centering
    \includegraphics[width=\columnwidth]{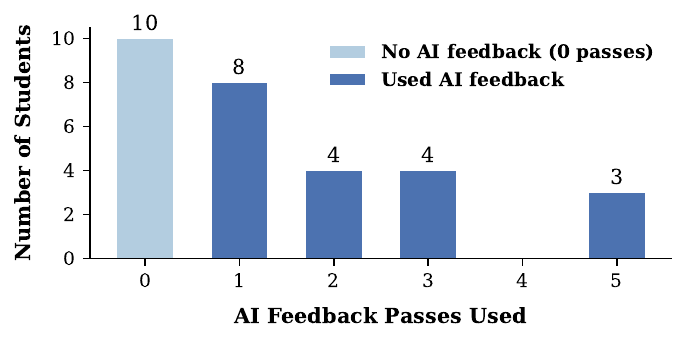}
    \caption{Distribution of AI feedback passes used per student.}
    \label{fig:passes}
\end{figure}

\subsection{Analysis}

Analyses were conducted in Python using \texttt{scipy.stats}. We report descriptive statistics for all key variables and the trend in mean AI composite scores across attempts (1--5). We use a Wilcoxon signed-rank test to compare first and final reviewed submissions among students with at least two review cycles, and a two-tailed Mann-Whitney U test to compare human final scores between AI-feedback users ($n=19$) and non-users ($n=8$). We also report Spearman correlations for focus ratio versus human final score and for passes used versus AI score gain, along with a Pearson correlation between AI last review score and human final score as a convergent validity check. Given the small sample ($N = 29$) and single-session, non-experimental design, all analyses are treated as exploratory, with effect sizes emphasized alongside $p$-values.

\section{Results}
\subsection{RQ1. Student Engagement Patterns}

Of the 29 students, 19 (65.5\%) used at least one AI review pass, while 10 (34.5\%) submitted without any AI feedback. Fig.~\ref{fig:passes} shows pass usage was right-skewed. One pass was most common ($n=8$), usage tapered to $n=4$ each at two and three passes, and no student stopped at exactly four. The distribution instead jumps directly to a smaller group that exhausted all five ($n=3$). This gap suggests two usage patterns, a light-touch majority sampling feedback once or twice, and a small minority treating all five passes as an exhaustive revision loop.
Students were highly focused. Mean focus ratio was $M = 0.824$ ($SD = 0.116$, range $0.56$--$1.00$), and mean task duration was $M = 77.3$ min ($SD = 5.0$). Competitive mechanics were used sparingly. Five students (17.2\%) attacked an opponent's pass and none shielded, suggesting most students prioritized writing over social interference even when the option was available.

\vspace{-1mm}
\subsection{RQ2. Writing Performance Across AI Feedback Attempts}
\begin{figure}[tbp]
    \centering
    \includegraphics[width=\columnwidth]{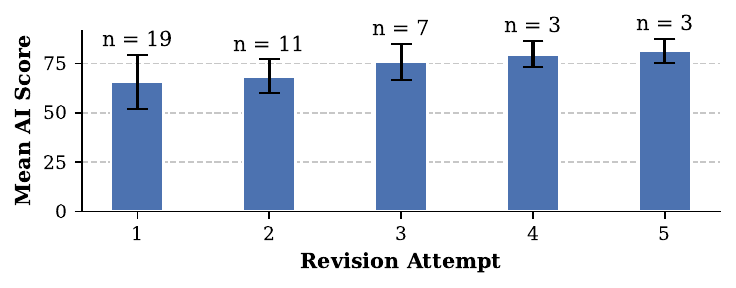}
    \caption{Mean AI composite scores (with standard deviation bars) by revision attempt.}
    \label{fig:scores}
\end{figure}

For students who used AI feedback ($n = 19$), mean \AIModel{} composite scores rose monotonically across attempts (Fig.~\ref{fig:scores}), climbing from $M \approx 65.6$ at Attempt~1 ($n=19$), through $\approx 68$ (Attempt~2, $n=11$) and $\approx 75$ (Attempt~3, $n=7$), to $\approx 79$--$81.3$ by Attempts~4--5 ($n=3$ each). The steepest gain falls between Attempts~2 and 3, and the shrinking error bars at later attempts reflect both fewer remaining students and less variable, higher-scoring drafts among those who persisted.
Mean AI score gain from first to last attempt was $M = 3.7$ ($SD = 7.4$, range $-5$ to $+22$, $n=19$). Among participants completing at least two review cycles ($n=11$), scores were significantly higher on the final than the first review, $W = 8.5$, $p = .032$ (one-tailed), with a large effect size (rank-biserial $= .74$). The three students who used all five cycles showed the largest gains, while a few students saw minor decreases between consecutive attempts, likely reflecting natural variation in AI scoring of near-equivalent drafts.

\vspace{-1mm}
\subsection{RQ3. AI Feedback, Human Scores, and Convergent Validity}

Human final scores for AI-feedback users ($n = 19$, $M = 76.9$) were descriptively higher than for non-users ($n = 8$, $M = 70.4$), but the difference was not significant (Mann-Whitney $U = 102.5$, $p = .167$, rank-biserial $= 0.349$), likely reflecting the small sample and the lack of random assignment. The Pearson correlation between students' last AI review score and their human final score was $r = 0.722$ ($p < .001$, $n = 19$), a strong convergence between \AIModel{} rubric-based scoring and the instructor's holistic evaluation, and a key finding supporting the validity of the AI feedback mechanism. Focus ratio, meanwhile, showed no significant association with human final score (Spearman $\rho = -0.147$, $p = .464$, $n=27$), while passes used showed a positive trend with AI score gain ($\rho = 0.428$, $p = .068$, $n=19$) that did not reach $\alpha=.05$, likely due to sample size. Students who attacked opponents ($M_{\text{human}} = 79.4$) scored about the same as those who did not ($M_{\text{human}} = 74.1$), suggesting attack behavior was not systematically tied to writing quality.

\begin{figure}[tbp]
    \centering
    \includegraphics[trim=0mm 4mm 0mm 0mm, clip, width=\columnwidth]{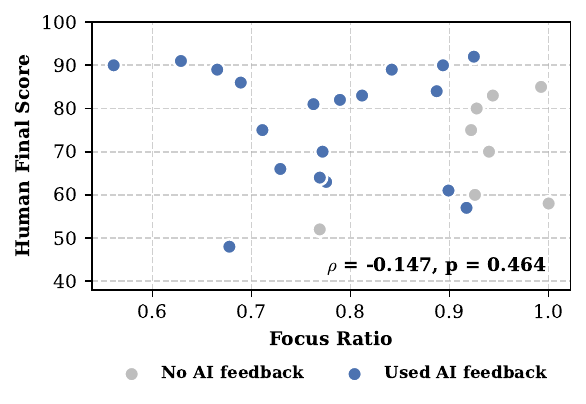}
    \caption{Scatter plot of focus ratio vs. human final score.}
    \label{fig:scatter}
\end{figure}

Table~\ref{tab:summary} juxtaposes the four analytical results. The AI-human correlation is far more reliable ($p < .001$) than any of the three engagement-related tests, none of which reach $\alpha=.05$, showing that convergent validity, not engagement, is this dataset's strongest signal. Fig.~\ref{fig:scatter} shows why. Focus ratio and human score do not co-vary systematically. Top scores ($\geq 90$) occur across the full focus range, including one student at focus $=0.56$, while several students above focus $0.9$ scored only in the low 60s. Non-users (grey) cluster at high focus ($0.85$--$1.0$) but span nearly the entire score range ($52$--$85$), so sustained on-screen attention alone did not guarantee a strong final essay.

\begin{table}[tbp]
\caption{Summary of Key Statistics}
\label{tab:summary}
\centering
\begin{tabular}{lcc}
\toprule
\textbf{Variable} & \textbf{$M$ ($SD$)} & \textbf{Range} \\
\midrule
Focus ratio              & $0.824$ $(0.116)$ & $0.56$--$1.00$ \\
Task duration (min)      & $77.3$ $(5.0)$    & -- \\
Passes used              & --                & $0$--$5$ \\
AI score gain            & $3.7$ $(7.4)$     & $-5$--$+22$ \\
Human final score        & $75.0$ $(13.1)$   & $48$--$92$ \\
\midrule
\multicolumn{3}{l}{\textit{Analytical results}} \\
\midrule
AI$_{\text{last}}$ vs.\ Human ($r$) & $0.722$*** & $p < .001$ \\
Passes used vs.\ Gain ($\rho$)      & $0.428$\textdagger & $p = .068$ \\
Focus ratio vs.\ Human ($\rho$)     & $-0.147$   & $p = .464$ \\
Users vs.\ Non-users ($U$)          & $102.5$    & $p = .167$ \\
\bottomrule
\multicolumn{3}{l}{\footnotesize \textdagger $p < .10$, *** $p < .001$}
\end{tabular}
\end{table}

\section{Discussion}
\subsection{AI Feedback as Evaluative Scaffolding}

The monotonic increase in AI-assessed scores across revision attempts, from $M = 65.6$ at Attempt~1 to $M = 81.3$ at Attempt~5 (Fig.~\ref{fig:scores}), indicates that students responded productively to the \AIModel{} scoring system. Viewed through a ZPD lens \cite{vygotsky1978mind}, the AI functions as a ``more knowledgeable other'' that externalizes performance standards, letting students close the gap between their draft and the rubric's expectations. This operationalizes the monitor--evaluate--adjust loop central to SRL theory \cite{zimmerman2000attaining} and aligns with multi-agent scaffolding frameworks showing that constraints on AI access prompt metacognitive monitoring and strategic resource allocation \cite{ryoo2025wip}.

A key qualification is that the system provided \textit{evaluative scaffolding}, scores and dimensional breakdowns, rather than fully elaborated formative feedback with explicit revision suggestions. Students had to infer what to revise from scores alone. This minimalist design helps prevent the technological dependency observed when students use fully generative AI tools \cite{park2025evaluating}, and the fact that scores nonetheless improved suggests that even minimal feedback, embedded in a high-stakes competitive frame, can motivate meaningful revision. Future iterations could progressively fade from explicit feedback toward scores-only as students approach their final revision, consistent with fading scaffolding in ZPD-informed instruction. The diminishing sample at higher attempts ($n = 19$ at Attempt~1 to $n = 3$ at Attempt~5, Fig.~\ref{fig:scores}) also reflects self-selection, since students who kept revising were likely more motivated or saw more improvement potential, so this survivorship pattern should temper how the later-attempt gains are interpreted.

\subsection{Gamification, Engagement, and Self-Determination}

The high mean focus ratio ($M = 0.824$) indicates the game context sustained behavioral engagement throughout the 60-minute window, consistent with Zhihao and Zhonggen \cite{zhihao2022impact}. Under SDT \cite{deci2000and}, the competitive narrative likely satisfied \textit{relatedness} (social competition) and \textit{relevance} (career framing), while the choice-based pass economy supported \textit{autonomy}. A theoretically interesting finding is the asymmetry between offensive mechanics (attack) and constructive mechanics (AI feedback). Despite comparable cost, only 5 students (17.2\%) attacked and none shielded, whereas 19 (65.5\%) used AI feedback (Fig.~\ref{fig:passes}). SDT's \textit{competence} need offers a parsimonious explanation, since students appeared to favor mechanics that improved their own ability over those that undermined others', consistent with Hajian et al.'s \cite{hajian2025motivational} argument that AI support and scaffolding construct learner motivation. The competitive frame appears to have generated motivational energy that students channeled into writing rather than social interference, a desirable outcome pedagogically. Future designs might tie competitive mechanics directly to writing quality (e.g., attack power scaling with AI score) so the two mechanics reinforce rather than compete.

\vspace{-1mm}
\subsection{AI-Human Score Convergence}

The strong correlation between AI last scores and human final scores ($r = 0.722$, $p < .001$) evidences convergent validity between \AIModel{}-based scoring and the instructor's holistic evaluation (Table~\ref{tab:summary}). This does not indicate perfect agreement, but supports AI scoring as a real-time formative tool within the game while human scoring remains appropriate for summative evaluation. The non-significant group difference between AI-feedback users and non-users ($p = .167$) should not be read as evidence of ineffectiveness. The study was not powered to detect causal effects, and the descriptive pattern (users $M = 76.9$ vs.\ non-users $M = 70.4$) is directionally consistent with a modest positive relationship.

\section{Conclusion}
This paper introduced \textit{The Brand War}, a gamified web application that embeds iterative \AIModel{}-powered writing feedback within a competitive narrative writing task for undergraduate EFL learners. An exploratory single-session study with $N = 29$ students in Taiwan found that nearly two-thirds of students engaged with the AI feedback system, AI-assessed scores rose monotonically across revision attempts, AI and human scores converged strongly ($r = 0.722$), and students maintained high task focus ($M = 82.4\%$) while preferring revision over competition even when competitive mechanics were available. These findings suggest that integrating AI evaluative scaffolding within a competitive game context is pedagogically feasible and motivationally coherent for EFL writing instruction, offering a practical deployment model that requires no extensive instructor training or real-time feedback provision. Limitations include the small, single-class sample without a control condition, which limits generalizability and precludes causal inference \cite{lin2021design}, the use of focus ratio as only a behavioral proxy for engagement, and the evaluative rather than fully formative nature of the AI feedback, whose inter-rater reliability against a larger essay corpus remains to be established. We would like to address these limitations in the future work.

\bibliographystyle{IEEEtran}
\bibliography{ref}

\end{document}